\documentclass{article}

     \usepackage[preprint]{nips_2018}

\usepackage[utf8]{inputenc} 
\usepackage[T1]{fontenc}    
\usepackage{hyperref}       
\usepackage{url}            
\usepackage{booktabs}       
\usepackage{amsfonts}       
\usepackage{nicefrac}       
\usepackage{microtype}      
\usepackage{graphicx} 
\usepackage{graphicx}
\usepackage{float}
\usepackage{subfig}

\usepackage{multirow}
\usepackage[table,xcdraw]{xcolor}

\title{When SAM Meets Medical Images: 
	An Investigation of Segment Anything Model (SAM) on Multi-phase Liver Tumor Segmentation}

\author{%
  Chuanfei Hu$^{1,3}$, Tianyi Xia$^2$, Shenghong Ju$^2$, Xinde Li$^{1,3}$ \\
  $^1$ School of Automation, Southeast University\\
  $^2$ Department of Radiology, Zhongda Hospital,  School of Medicine, Southeast University\\
  Nanjing, China \\
  $^3$ Nanjing Center for Applied Mathematics, Nanjing, China
}

\begin{document}

\maketitle

\begin{abstract}
  Learning to segmentation without large-scale samples is an inherent capability of human.
  Recently, Segment Anything Model (SAM) performs the significant zero-shot image segmentation,
  attracting considerable attention from the computer vision community.
  Here, we investigate the capability of SAM for medical image analysis, especially for multi-phase liver tumor segmentation (MPLiTS),
  in terms of prompts, data resolution, phases.
  Experimental results demonstrate that there might be a large gap between SAM and expected performance.
  Fortunately, the qualitative results show that SAM is a powerful annotation tool for the community of interactive medical image segmentation.
\end{abstract}

\section{Introduction}

Large language models (LLMs), such ChatGPT\footnote{https://help.openai.com/en/articles/6825453-chatgpt-release-notes}, have performed the superior capability on various natural language processing (NLP) tasks. 
Since the friendly application program interface (API) of LLMs are accessible, 
the numerous amazing applications have been conducted, such as ChatGPT+Midjourney, HuggingGPT,
in sense that the large models~\cite{bommasani2021opportunities} would be a powerful tool to change the paradigm of our workflow.

More recently, Segment Anything Model (SAM)~\cite{kirillov2023segment} is proposed, which is trained on a novel large visual dataset with 1 billion segmentation masks, termed as SA-1B.
It might be one of the most promising large models to handle the foundation tasks towards various scenarios for the computer vision community.
Meanwhile, due to the capability of zero-shot image segmentation with some prompts, 
SAM is attractive particularly for the medical image analysis where the annotations and samples are rare and laborsome.

Motivated by~\citet{deng2023segment, zhang2023attack, han2023segment, zhang2023comprehensive, zhang2023asurvey}, we investigate the performance of SAM on multi-phase liver tumor segmentation (MPLiTS).
SAM with three variants (ViT-B, ViT-L, ViT-H) is validated in terms of prompts, data resolution, phases on in-house dataset.
The results reveal that SAM might not achieve the expected performance for MPLiTS.
However, the fact without doubt is that SAM is a powerful tool to boost the efficiency of annotation significantly.

\section{Experiments}

In this section, we first introduce the in-house MPLiTS dataset. 
Then, the evaluation protocols and implementation details are described,
while the quantitative and qualitative results are reported.

\subsection{Dataset}
The in-house dataset collects 388 patients 
with 1552 multi-phase contrast-enhanced computed tomography (CECT) volumes. 
All volumes are acquired by Philips iCT 256 scanners with non-contrast, arterial, portal venous, and delayed phases. 
The in-plane size of volumes is $512 \times 512$ with spacing ranges from 0.560 mm to 0.847 mm, 
and the number of slices ranges from 25 to 89 with spacing 3.0 mm.
The volumes of four phases are co-registered into venous phase by Elastix~\cite{Klein2010elastix} toolbox.
The ground truths of all volumes are annotated by two radiologists (with 10 years and 20 years of experiences in liver imaging, respectively),
where the hepatocellular carcinoma (HCC) lesions are outlined in the delayed phase with reference to the other phases.

\subsection{Evaluation Protocols}
The multi-phase CECT volumes of in-house dataset are divided into five folds with the same ratio in terms of patient, 
where the average performance of SAM for the five-fold parts is employed. 
Dice global score (DGS) is utilized to evaluate the performance of liver tumor segmentation,
which can be denoted as follows:
\begin{equation} 
	{DGS}_{s}= \frac{1}{N} \sum_{i=1}^{N} \frac{2|X^{s}_{i} \cap Y_{i}|}{|X^{s}_{i}|+|Y_{i}|}
\end{equation}
where $X^{s}_{i} \in \mathbb{R}^{H \times W}$ is a CECT image of the $s$-th phase from a set $\mathcal{X}_{i} = \{ {X}_{i}^{s} | {s \in \mathcal{S} \}}$ 
with a set of multi-phase $\mathcal{S} = \{${non-contrast (NC)}, {arterial (ART)}, {portal venous (PV)}, {delayed (DE)}$\}$.
$Y_{i}$ is the ground truth mask, $N$ is the number of volume slices in our dataset. $H$ and $W$ are the resolution of the CECT image.


\subsection{Implementation Details}
The experiments are conducted on a work station with NVIDIA Tesla A100 GPUs. 
The variants of SAM with ViT-B, ViT-L and ViT-H are conducted separately to segment the CECT images,
while the different validation settings $\mathcal{T}^{\mathcal{M}}_{\mathcal{P}, \mathcal{R}}$ are introduced in terms of prompts $\mathcal{P}$, data resolution $\mathcal{R}$, phases $\mathcal{M}$.
Since the outputs of SAM are multiply, we select one of the superior tumor masks of SAM for multi-phase as the results of $\mathcal{T}^{\mathcal{M}}_{\mathcal{P}, \mathcal{R}}$.

Specifically, 
\begin{itemize}
\item $\mathcal{P}$, we select the point mode of prompts with the various numbers $\mathcal{P} = \{$1, 5, 10, 20$\}$.
\item $\mathcal{R}$, the resolutions of CECT images are selected with $\mathcal{R} = \{$224, 512, 1024$\}$.
\item $\mathcal{M}$, since SAM is not proposed to multi-phase input data, two modes are designed to aggregate the multi-phase results. 
$\mathcal{M}=avg$ and $\mathcal{M}=max$ denote the multi-phase results are aggregated via average and maximum operations, respectively.
\end{itemize}

\subsection{Results}

\begin{figure*}[t]
	\centering
	\subfloat[CECT images with four phases including NC, ART, PV, DE.]{
		\includegraphics[width=0.85\linewidth]{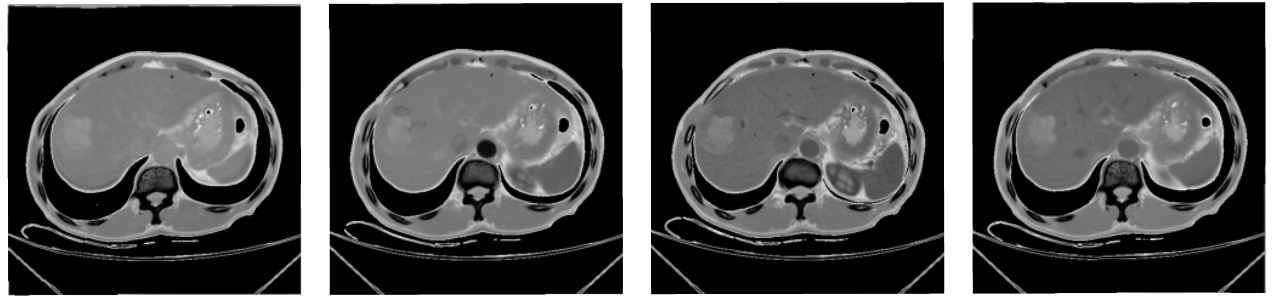}
	}
	
	\vspace{-3mm}
	\subfloat[The superior results of $\mathcal{T}^{max}_{1, 1024}$, $\mathcal{T}^{max}_{5, 1024}$, 
	and $\mathcal{T}^{max}_{20, 1024}$ are 0.4668, 0.7923, and 0.8761, respectively.
	The last image is the ground truth.
	]{
		\includegraphics[width=0.85\linewidth]{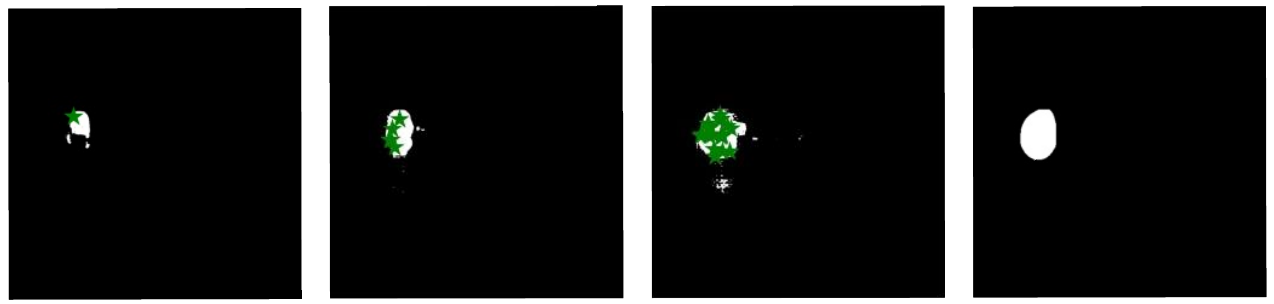}
	}
	\caption{Visual example of results with the various $\mathcal{P}$.}
	\label{fig:prompt}
\end{figure*}

\begin{figure*}[t]
	\centering
	\subfloat[CECT images with four phases including NC, ART, PV, DE.]{
		\includegraphics[width=0.85\linewidth]{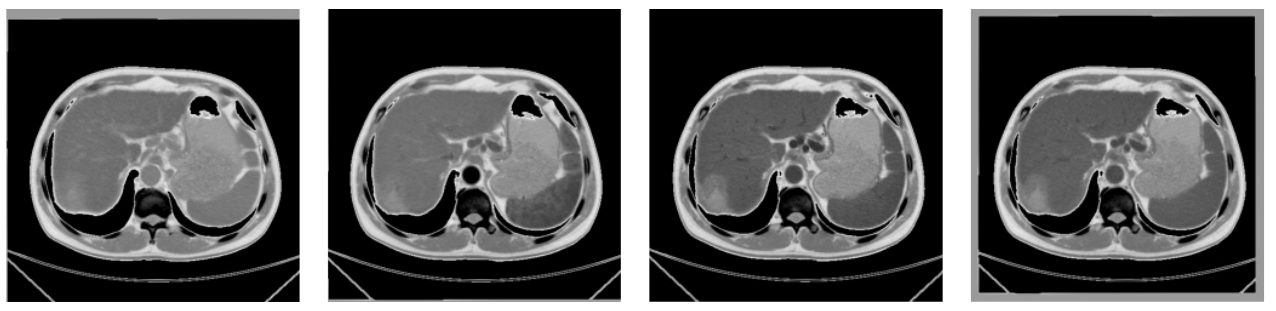}
	}
	
	\vspace{-3mm}
	\subfloat[The superior results of $\mathcal{T}^{max}_{10, 224}$, $\mathcal{T}^{max}_{10, 512}$, 
	and $\mathcal{T}^{max}_{10, 1024}$ are 0.8397, 0.6576, and 0.5449, respectively.
	The last image is the ground truth.
	]{
		\includegraphics[width=0.85\linewidth]{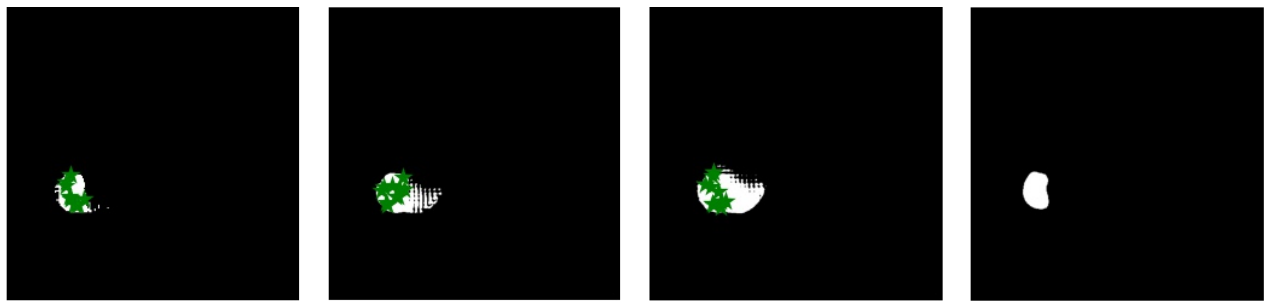}
	}
	\caption{Visual example of results with the various $\mathcal{R}$.}
	\label{fig:resolution}
\end{figure*}

\begin{figure*}[t]
	\centering
	\subfloat[CECT images with four phases including NC, ART, PV, DE.]{
		\includegraphics[width=0.85\linewidth]{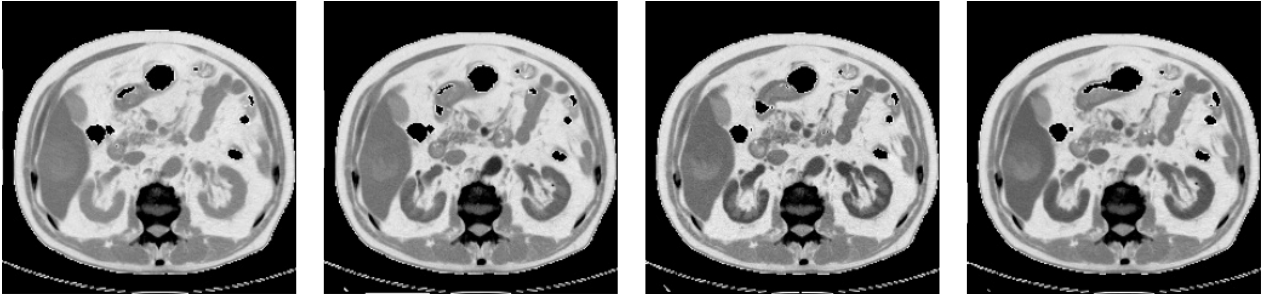}
	}
	
	\vspace{-3mm}
	\subfloat[The superior results of $\mathcal{T}^{max}_{20, 224}$ with ViT-b, ViT-l, and ViT-h are 0.4035, 0.8920, and 0.9246, respectively.
	The last image is the ground truth.
	]{
		\includegraphics[width=0.85\linewidth]{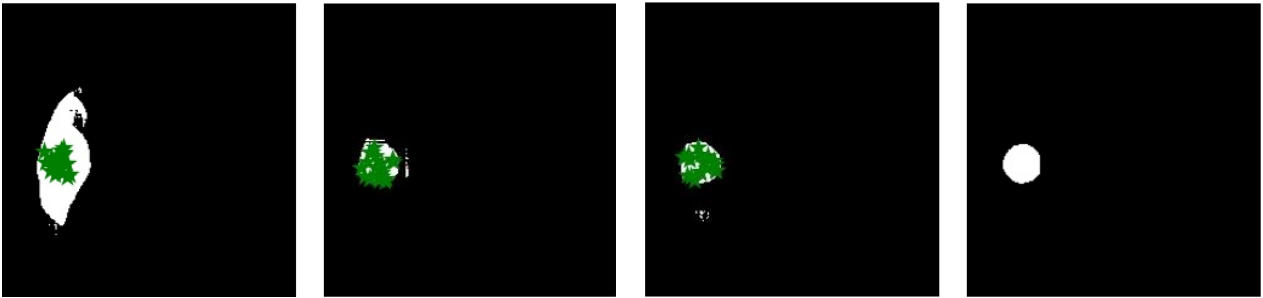}
	}
	\caption{Visual example of results with the various $\mathcal{R}$.}
	\label{fig:backbone}
\end{figure*}

The experimental results are reported in Table~\ref{tab:total_res},
while there are three observations can be summarized as follows:
\begin{itemize}
	\item \textit{The number of prompt points $\mathcal{P}$ dominates the performance of SAM.}  
	The tendency of the performance becomes superior with incremental $\mathcal{P}$. (Figure~\ref{fig:prompt})
	\item \textit{The larger resolution of data might not be the better.} 
	The increasing $\mathcal{R}$ of CECT images do not improve the overall performance of SAM. (Figure~\ref{fig:resolution})
	\item \textit{The incremental $\mathcal{P}$ would activate the advantage of ``stronger'' encoder.} (Figure~\ref{fig:backbone})
\end{itemize}

\begin{figure*}[t]
	\centering
	\includegraphics[width=0.85\linewidth]{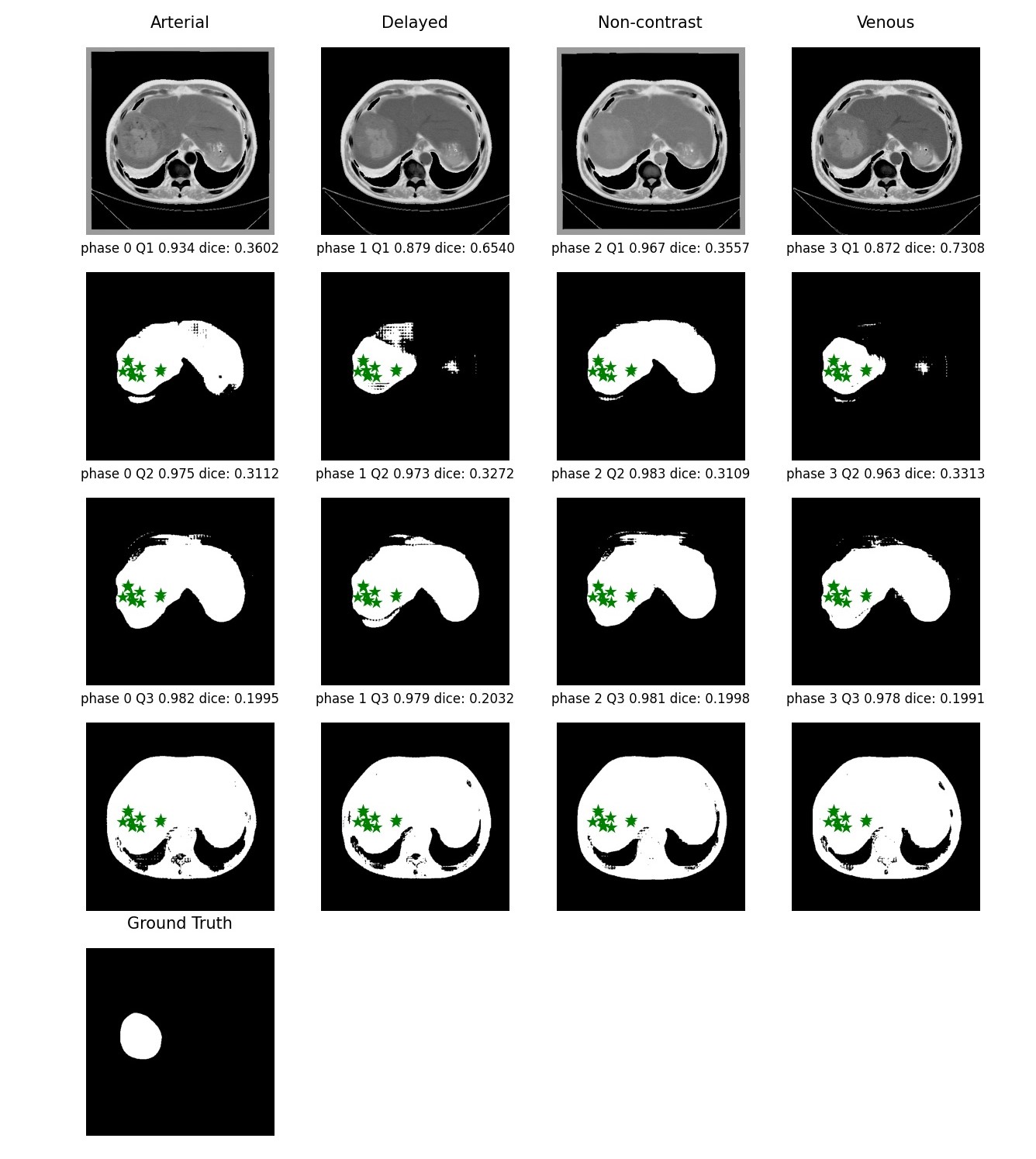}
	\caption{Visual example of the entire result with $\mathcal{T}^{\mathcal{M}}_{10, 512}$.}
	\label{fig:entire}
\end{figure*}

However, compared with U-Net~\cite{ronneberger2015u}, there is still a large gap when SAM with the few prompt points.
Fortunately, SAM is seen like a powerful annotation tool with enough human guidance (with $\mathcal{P} = $ 20).
The example of the entire result is shown in Figure~\ref{fig:entire}.

\begin{table}[]
	\caption{Validation of SAM with the different settings $\mathcal{T}^{\mathcal{M}}_{\mathcal{P}, \mathcal{R}}$ in terms of DCS (\%).}
	\label{tab:total_res}
	\centering
	\resizebox{0.5\linewidth}{!}{
	\begin{tabular}{c|cc|ccc|c}
		\toprule
		$\mathcal{P}$                   & $\mathcal{R}$                     & $\mathcal{M}$          & ViT-b    & ViT-l        & ViT-h  & U-Net   \\
		\midrule
		&                        &  224  & 0.2569   &  0.4033  & 0.4000 & \textbf{0.7932}\\
		&                        &  512   &   0.2496   &  0.3511  &  0.3431 & \textbf{0.7947} \\
		& \multirow{-3}{*}{ $avg$} &  1024 & 0.2507  & 0.3572 & 0.3419 & \textbf{0.8061}\\ \cline{2-7}
		&&&&&\\[-3mm]  
		&                       & 224   &   0.4165 &   0.6308 &  0.6153 \\
		&                       & 512  &  0.3780 &  0.5556 & 0.5183 \\
		\multirow{-6}{*}{1} & \multirow{-3}{*}{$max$} & 1024 &  0.3877  & 0.5616& 0.5184\\
		
		\midrule
		&                        &  224  & 0.2818  & 0.4197 &  0.4828\\
		&                        &  512   &   0.2656 &   0.3662   &  0.4004 \\
		& \multirow{-3}{*}{$avg$} &  1024 & 0.2687  & 0.3684 & 0.3978  \\ \cline{2-7}
		&&&&&\\[-3mm] 
		&                       & 224 &   0.4406 &     0.6395 & 0.6982 \\
		&                       & 512  &  0.4008 & 0.5802 & 0.5915 \\
		\multirow{-6}{*}{5} & \multirow{-3}{*}{$max$} & 1024  &   0.4017   & 0.5823 &   0.5853\\

		\midrule
		&                        &  224  & 0.3076  & 0.4360 & 0.5575 \\ 
		&                        &  512    &  0.2783 & 0.3870 &  0.4738 \\
		& \multirow{-3}{*}{$avg$} &  1024 & 0.2826& 0.3894 & 0.4748 \\\cline{2-7}
		&&&&&\\[-3mm]
		&                       & 224   &  0.4740 &   0.6451 &  0.7431 \\
		&                       & 512  & 0.4142 & 0.5952 &  0.6651\\
		\multirow{-6}{*}{10} & \multirow{-3}{*}{$max$} & 1024  & 0.4190  &  0.5978 &  0.6697\\
		
		\midrule 
		&                        &  224  & 0.3367   & 0.4307 &   0.6109 \\ 
		&                        &  512   & 0.2995    &  0.3860  &  0.5399 \\
		& \multirow{-3}{*}{$avg$} &  1024 & 0.3017  &  0.3881 & 0.5359 \\\cline{2-7}
		&&&&&\\[-3mm]
		&                       & 224  &  0.5062 & 0.6295 &  0.7630\\
		&                       & 512  & 0.4396 &  0.5834 &  0.7086\\
		\multirow{-6}{*}{20} & \multirow{-3}{*}{$max$} & 1024  & 0.4447&    0.5866 &  0.7025 \\
		\bottomrule
	\end{tabular}
}
\end{table}

\section{Conclusion}
In this paper, we conducted a preliminary investigation of SAM for medical image analysis, 
especially for multi-phase liver tumor segmentation (MPLiTS), in terms of prompts, data resolution, phases.
The experimental results reveal the superior capability of SAM as an annotation tool, and the room for improvement in MPLiTS.
A further investigation would be conducted in terms of the comprehensively aspects,
which could be a guidance to the community of MPLiTS.

\bibliographystyle{plainnat}
\bibliography{aaai22.bib}

\begin{thebibliography}{9}
\providecommand{\natexlab}[1]{#1}
\providecommand{\url}[1]{\texttt{#1}}
\expandafter\ifx\csname urlstyle\endcsname\relax
  \providecommand{\doi}[1]{doi: #1}\else
  \providecommand{\doi}{doi: \begingroup \urlstyle{rm}\Url}\fi

\bibitem[Bommasani et~al.(2021)Bommasani, Hudson, Adeli, Altman, Arora, von
  Arx, Bernstein, Bohg, Bosselut, Brunskill,
  et~al.]{bommasani2021opportunities}
Rishi Bommasani, Drew~A Hudson, Ehsan Adeli, Russ Altman, Simran Arora, Sydney
  von Arx, Michael~S Bernstein, Jeannette Bohg, Antoine Bosselut, Emma
  Brunskill, et~al.
\newblock On the opportunities and risks of foundation models.
\newblock \emph{arXiv preprint arXiv:2108.07258}, 2021.

\bibitem[Deng et~al.(2023)Deng, Cui, Liu, Yao, Remedios, Bao, Landman, Wheless,
  Coburn, Wilson, et~al.]{deng2023segment}
Ruining Deng, Can Cui, Quan Liu, Tianyuan Yao, Lucas~W Remedios, Shunxing Bao,
  Bennett~A Landman, Lee~E Wheless, Lori~A Coburn, Keith~T Wilson, et~al.
\newblock Segment anything model (sam) for digital pathology: Assess zero-shot
  segmentation on whole slide imaging.
\newblock \emph{arXiv preprint arXiv:2304.04155}, 2023.

\bibitem[Han et~al.(2023)Han, Zhang, Qiao, Qamar, Jung, Lee, Bae, and
  Hong]{han2023segment}
Dongsheng Han, Chaoning Zhang, Yu~Qiao, Maryam Qamar, Yuna Jung, SeungKyu Lee,
  Sung-Ho Bae, and Choong~Seon Hong.
\newblock Segment anything model (sam) meets glass: Mirror and transparent
  objects cannot be easily detected.
\newblock \emph{arXiv preprint arXiv:2305.00278}, 2023.

\bibitem[Kirillov et~al.(2023)Kirillov, Mintun, Ravi, Mao, Rolland, Gustafson,
  Xiao, Whitehead, Berg, Lo, et~al.]{kirillov2023segment}
Alexander Kirillov, Eric Mintun, Nikhila Ravi, Hanzi Mao, Chloe Rolland, Laura
  Gustafson, Tete Xiao, Spencer Whitehead, Alexander~C Berg, Wan-Yen Lo, et~al.
\newblock Segment anything.
\newblock \emph{arXiv preprint arXiv:2304.02643}, 2023.

\bibitem[Klein et~al.(2010)Klein, Staring, Murphy, Viergever, and
  Pluim]{Klein2010elastix}
Stefan Klein, Marius Staring, Keelin Murphy, Max~A. Viergever, and Josien P.~W.
  Pluim.
\newblock elastix: A toolbox for intensity-based medical image registration.
\newblock \emph{IEEE Transactions on Medical Imaging}, 29\penalty0
  (1):\penalty0 196--205, 2010.
\newblock \doi{10.1109/TMI.2009.2035616}.

\bibitem[Ronneberger et~al.(2015)Ronneberger, Fischer, and
  Brox]{ronneberger2015u}
Olaf Ronneberger, Philipp Fischer, and Thomas Brox.
\newblock U-net: Convolutional networks for biomedical image segmentation.
\newblock In \emph{Medical Image Computing and Computer-Assisted Intervention
  -- MICCAI 2015}, pages 234--241. Springer, 2015.
\newblock \doi{10.1007/978-3-319-24574-4\_28}.

\bibitem[Zhang et~al.(2023{\natexlab{a}})Zhang, Zheng, Li, Qiao, Kang, Shan,
  Zhang, Qin, Rameau, Bae, et~al.]{zhang2023asurvey}
Chaoning Zhang, Sheng Zheng, Chenghao Li, Yu~Qiao, Taegoo Kang, Xinru Shan,
  Chenshuang Zhang, Caiyan Qin, Francois Rameau, Sung-Ho Bae, et~al.
\newblock A survey on segment anything model (sam): Vision foundation model
  meets prompt engineering.
\newblock 2023{\natexlab{a}}.

\bibitem[Zhang et~al.(2023{\natexlab{b}})Zhang, Zhang, Kang, Kim, Bae, and
  Kweon]{zhang2023attack}
Chenshuang Zhang, Chaoning Zhang, Taegoo Kang, Donghun Kim, Sung-Ho Bae, and
  In~So Kweon.
\newblock Attack-sam: Towards evaluating adversarial robustness of segment
  anything model.
\newblock \emph{arXiv preprint arXiv:2305.00866}, 2023{\natexlab{b}}.

\bibitem[Zhang et~al.(2023{\natexlab{c}})Zhang, Liu, Cui, Huang, Lin, Yang, and
  Hu]{zhang2023comprehensive}
Chunhui Zhang, Li~Liu, Yawen Cui, Guanjie Huang, Weilin Lin, Yiqian Yang, and
  Yuehong Hu.
\newblock A comprehensive survey on segment anything model for vision and
  beyond.
\newblock \emph{arXiv preprint arXiv:2305.08196}, 2023{\natexlab{c}}.

\end{thebibliography}

\end{document}